\documentclass[12pt,draftclsnofoot,onecolumn]{IEEEtran}

\usepackage{amssymb}
\usepackage{amsmath}
\usepackage[dvips]{graphicx}
\usepackage[caption=false]{subfig}
\usepackage{kbordermatrix}

\newtheorem{definition}{Definition}

\renewcommand{\kbldelim}{(}
\renewcommand{\kbrdelim}{)}

\title{Cognitive MAC Protocols Using Memory \\for Distributed Spectrum Sharing \\Under Limited Spectrum Sensing}
\author{Jaeok Park and Mihaela van der Schaar\thanks{The authors are with Electrical Engineering Department, University of
California, Los Angeles (UCLA), 420 Westwood Plaza,
Los Angeles, CA 90095-1594, USA. e-mail: \{jaeok, mihaela\}@ee.ucla.edu.}}
\date{}

\begin{document}

\maketitle

\begin{abstract}

The main challenges of cognitive radio include spectrum sensing at the physical (PHY)
layer to detect the activity of primary users and spectrum
sharing at the medium access control (MAC) layer to coordinate access among
coexisting secondary users.
In this paper, we consider a cognitive radio network in which a primary user shares
a channel with secondary users that cannot distinguish the signals of the primary user
from those of a secondary user.
We propose a class of distributed cognitive MAC protocols to achieve efficient spectrum sharing
among the secondary users while protecting the primary user from potential interference
by the secondary users.
By using a MAC protocol with one-slot memory,
we can obtain high channel utilization by the secondary users while limiting interference to
the primary user at a low level.
The results of this paper suggest the possibility of utilizing MAC design in cognitive radio networks
to overcome limitations in spectrum sensing at the PHY layer as well as to achieve spectrum sharing
at the MAC layer.

\end{abstract}


\section{Introduction}

Today's expanding demand for wireless services has necessitated
cognitive radio technology in order to overcome the limitations of the conventional
static spectrum allocation policy. Cognitive radio technology enables a
more efficient use of limited spectrum resources by allowing unlicensed
users (or secondary users) to opportunistically utilize licensed spectral bands.
The main challenges of cognitive radio include \emph{spectrum sensing} at the physical (PHY)
layer to detect the activity of licensed users (or primary users) and \emph{spectrum
sharing} at the medium access control (MAC) layer to coordinate access among
coexisting secondary users \cite{akyildiz}. Spectrum sensing is needed to
identify spectrum opportunities or spectrum holes, while spectrum sharing helps secondary users
achieve an efficient and fair use of identified spectrum opportunities.

In this paper, we study a MAC protocol design problem for a cognitive radio
network in which a primary user shares a spectral band (or a channel) with
multiple secondary users. One of the main assumptions of our model is that
the secondary users have limited spectrum sensing capability at the PHY layer in the sense that
they are unable to distinguish between the activities (i.e., spectrum access) of
the primary user and a secondary user.
In other words, the secondary users can sense whether the channel is idle or busy,
but when the channel is sensed busy, they do not know whether the channel is accessed
by the primary user or not.
This assumption contrasts with and is weaker than the prevailing assumption, made
in previous work on MAC design for cognitive radio, that sensing at the PHY layer is perfect
in that secondary users can always detect the presence of primary users (see, for example, \cite{chou},\cite{fattahi}).
\cite{zhao} relaxes the assumption of perfect spectrum sensing and considers sensing
errors at the PHY layer. However,
\cite{zhao} requires that the signals of primary users be statistically distinguishable from
those of secondary users. On the contrary, our assumption is valid when the signals of
primary users are (statistically) indistinguishable from
those of secondary users.

Another key assumption we maintain is that explicit coordination messages cannot be communicated
between a central controller and a user, or between users.
This implies that the primary user cannot broadcast its presence to the secondary users
for spectrum sensing
and that centralized scheduling schemes such as TDMA cannot be used
for spectrum sharing. Again, this assumption contrasts with and is weaker than
the assumption made in existing work that requires central controllers or
dedicated control channels (see, for example, \cite{chou},\cite{fattahi}). As pointed out in
\cite{akyildiz}, in cognitive radio networks, protocols requiring broadcast
messages cause a major problem due to the lack of a reliable control channel
as a channel has to be vacated whenever a primary user returns to the channel.

Our protocol design for the secondary users is based on MAC protocols
with memory, which are formally presented in \cite{park}. Under a
protocol with memory, users adjust their transmission parameters depending
on the local histories of their own transmission actions and feedback
information. Hence, protocols with memory can be implemented in a distributed way without
explicit message passing for any given sensing ability of users.
Moreover, by exploiting information embedded in local histories,
protocols with memory enable a secondary user to ``change its
transmitter parameters based on interaction with the environment
in which it operates,'' as demanded by the definition of cognitive radio \cite{FCC}.

In \cite{park}, we have focused on the problem of achieving coordinated access among
symmetric users by using a protocol with memory.
In a cognitive radio network, where a primary user exists, another kind of coordination is
needed to ensure that the secondary users do not interfere with the primary user.
In this paper, we show that a class of protocols with one-slot memory can achieve high
channel utilization by the secondary users while protecting the primary user
at a desired level. We also show that the system performance can be improved by utilizing
longer memory.
The results of this paper suggest that a carefully
designed MAC protocol can be used in place of an algorithm for primary user detection
at the PHY layer. The main contribution of this paper is to illustrate
the possibility of utilizing MAC design to overcome limitations in
spectrum sensing at the PHY layer as well as to achieve spectrum sharing
at the MAC layer.

In recent years, there have been burgeoning research efforts involving
cognitive radio networks. Due to space limitations, we review only a few of
them, focusing on the most related work, and refer the interested reader to
\cite{akyildiz} for a comprehensive survey. \cite{chou} examines
gains from spectrum agility in terms of spectrum utilization. Our model
corresponds to the non-agile case of \cite{chou} as secondary users in
our model stay in the same channel for the considered horizon of time.
This is because our model is not equipped with ideal control devices
as assumed in \cite{chou}. \cite{fattahi} uses a mechanism design
approach to determine the allocation of spectrum opportunities to selfish secondary
users. \cite{zhao} analyzes the decision of secondary users to sense
and access channels using a partially observable Markov decision process
framework. \cite{shuang} evaluates performance under two spectrum access
schemes using different sensing, back-off, and transmission mechanisms.
\cite{hkim} develops a sensing-period optimization mechanism and an optimal
channel-sequencing algorithm for efficient discovery of spectrum opportunities.
\cite{law} models the interactions between secondary users as a non-cooperative
game and derives the price of anarchy. A survey on MAC protocols for cognitive
radio networks is presented in \cite{wang}--\cite{cormio}.

The rest of this paper is organized as follows. In Section II, we describe
our system model. In Section III, we formulate MAC protocols,
performance metrics, and a protocol design problem. In Section IV, we explain how to compute the
performance metrics for a given protocol, using Markov chains. In Section V,
we solve the protocol design problem numerically.
In Section VI, we discuss how the proposed protocols can be enhanced
by utilizing longer memory.
In Section VII, we conclude this paper.

\section{System Model}

We consider a licensed channel in a slotted Aloha-type network, as in \cite{park} and \cite{ma},
with a single primary user and $N$ secondary users. We assume that $N$
is fixed over time. Time is divided into slots
of equal length,
and the primary and secondary users maintain synchronized time slots.
A user can attempt to transmit a packet or wait in a slot in which it has a packet to transmit.
Due to interference, only one user can transmit successfully in a slot, and
simultaneous transmission by more than one user results in a collision.

The traffic of the primary user arrives following a stochastic process.
We assume that an arrival of traffic generates multiple packets, the
average number of which is denoted by $T_{pac}$, and that
the average time interval (measured in slots) between two consecutive arrivals of traffic,
denoted by $T_{int}$, is larger than $T_{pac}$.\footnote{A scenario that fits into our assumptions
is one where the primary user has bursty traffic.}
In each slot, the primary user has either a packet to transmit or none depending on
traffic arrivals and transmission results.
The state of the primary user, denoted by $y_p$, is said to be \emph{on} if the primary user
has a packet to transmit and \emph{off} otherwise. A similar \emph{on}-\emph{off} model for
the primary user can be found in \cite{chou} and \cite{hkim}.\footnote{Under
perfect sensing assumed in \cite{chou} and \cite{hkim}, the
duration of \emph{on} and \emph{off} periods is independent of the existence
of the secondary users because the secondary users can be required to back off when
they sense the activity of the primary user.
On the contrary, under limited sensing in our model, an \emph{on}
period becomes longer while an \emph{off} period becomes shorter as the secondary users
create more collisions with the primary user. This fact is
taken into account in the objective of the protocol design problem formulated in Section III.}

Each secondary user always has packets to transmit.
After a user makes a transmission attempt, it learns whether the transmission is
successful or not using an acknowledgement (ACK) response.
The secondary users have the sensing ability to find out whether the channel is accessed
or not while they wait. However, when the channel is sensed busy, they do not
obtain information about whether the primary user accessed the channel or not.
This assumption limits the ability of the secondary users to detect the presence
of the primary user. Using the information from ACK responses and sensing,
a secondary user can classify a slot into the four states, \emph{idle}, \emph{busy}, \emph{success}, and \emph{failure},
as in \cite{hamed}. The state of secondary user $i$, denoted by $y_i$,
is \emph{idle} if no user transmits, \emph{busy} if secondary user $i$
does not transmit but at least one other user transmits, \emph{success} if
secondary user $i$ transmits and succeeds, and \emph{failure} if secondary user $i$ transmits
but fails.

\section{Protocol Description and Problem Formulation}

\subsection{Protocol Description}

\subsubsection{Protocol for the Primary User}

The decision rule for the primary user is to transmit
whenever it has a packet to transmit. Note that the primary user
does not need to modify its decision rule for coexistence with
the secondary users, which is consistent with the requirements of
cognitive radio networks.

\subsubsection{Protocol for the Secondary Users}

The decision rule for the secondary users is prescribed by a protocol with one-slot memory \cite{park}.
A protocol with one-slot memory specifies a transmission
probability for each possible state of the previous slot, and thus it can be
formally represented by a function $f:\mathcal{Y}_{s} \rightarrow [0,1]$,
where $\mathcal{Y}_{s}$ is the set of the states of a secondary user, i.e.,
$\mathcal{Y}_{s} = \{\textrm{\emph{idle}}, \textrm{\emph{busy}}, \textrm{\emph{success}}, \textrm{\emph{failure}} \}$.
A secondary user whose state is $y \in \mathcal{Y}_{s}$ in the previous slot
transmits with probability $f(y)$ in the current slot. We provide two definitions
about the properties of a protocol with one-slot memory.

\begin{definition}
A protocol $f$ with one-slot memory is \emph{non-intrusive} if $f(\textrm{\emph{busy}}) = 0$.
\end{definition}

When the secondary users follow a non-intrusive protocol, they wait in a slot
following a busy slot.
Thus, a non-intrusive protocol allows the primary user not to be interrupted
by the secondary users once it has a successful transmission.

\begin{definition}
A protocol $f$ with one-slot memory has the \emph{fairness level} $\theta \in (0,1]$
if the average number of consecutive successes by a secondary user while the primary
user does not transmit is $1/\theta$, or
\begin{align} \label{eq:fair}
1 - f(\textrm{\emph{success}}) (1-f(\textrm{\emph{busy}}))^{N-1} = \theta.
\end{align}
\end{definition}

Suppose that there is no transmission by the primary user. Once a secondary user succeeds, it has a successful transmission in the next
slot with probability $f(\textrm{\emph{success}}) (1-f(\textrm{\emph{busy}}))^{N-1}$, and thus the average
number of consecutive successes is given by $1/[1-f(\textrm{\emph{success}}) (1-f(\textrm{\emph{busy}}))^{N-1}]$.
As the fairness level is smaller, a secondary user keeps using
the channel for a longer period once it succeeds, which makes other secondary users wait longer
until they have a successful transmission. In \cite{ma}, a protocol with fairness level $\theta$
is said to be $M$-short-term fair if $1/\theta \leq M$.

\subsection{Performance Metrics}

\subsubsection{Collision Probability of the Primary User}

In overlay spectrum sharing, it is important to protect the
primary user from interruption by the secondary users. We measure
interference experienced by the primary user by the collision probability
of the primary user, defined as
\begin{align*}
P_c = \frac{\textrm{No. of collisions experienced
by PU}}{\textrm{No. of transmission attempts by PU}},
\end{align*}
where PU represents ``primary user.'' That is, the collision probability
of the primary user is the probability that it experiences
a collision when it attempts to transmit a packet.

\subsubsection{Channel Utilization of the Secondary Users}

We measure the utilization of spectrum opportunities by
the success probability of the secondary users, defined as
\begin{align*}
P_s = \frac{\textrm{No. of successes by SUs}}{\textrm{No. of slots in which PU is \emph{off}}},
\end{align*}
where SU represents ``secondary user.'' In other words, the success probability of the secondary users
is the probability that a secondary user has a successful transmission
when the primary user has no packet to transmit.
The channel utilization (or throughput) of the secondary users is defined as the proportion of
time slots in which a secondary user has a successful transmission, i.e.,
\begin{align*}
C_s = \frac{\textrm{No. of successes by SUs}}{\textrm{No. of slots}}.
\end{align*}

\subsubsection{Channel Utilization of the System}

The channel utilization of the system is defined as the proportion of
time slots in which a successful transmission occurs, i.e.,
\begin{align*}
C = \frac{\textrm{No. of successes}}{\textrm{No. of slots}}.
\end{align*}

\subsubsection{Computation of the Performance Metrics and Performance Bounds}

We define an \emph{on} period and an \emph{off} period as a period
in which the state of the primary user is \emph{on} and \emph{off}, respectively,
between two consecutive arrivals of traffic.
Let $T_{\textrm{\emph{on}}}$ and $T_{\textrm{\emph{off}}}$ be the average length
(measured in slots) of an \emph{on} period and an \emph{off} period, respectively.
Then the average time interval between two consecutive arrivals of traffic can be decomposed as
$T_{int} = T_{\textrm{\emph{on}}} + T_{\textrm{\emph{off}}}$.
Let $T_{col}$ be the average number of collisions that the primary user experiences
while transmitting packets generated by an arrival of traffic.
We assume that $T_{col} < T_{int} - T_{pac}$ to assure the stability of the system.
Since the primary user transmits whenever it has a packet to transmit, it has either
a successful transmission or a collision when its state is \emph{on}.
Hence, an \emph{on} period can be decomposed into slots in which the primary user succeeds
and those in which it collides, i.e., $T_{\textrm{\emph{on}}} = T_{pac} + T_{col}$.
Let $T_s$ and $T_{\textrm{\emph{ns}}}$ be the average numbers of slots
in which one and none, respectively, of the
secondary users has a successful transmission between two consecutive arrivals of traffic.
Given the protocol for the primary user and our contention model,
a secondary user can have a successful transmission only when the state of the
primary user is \emph{off}. Thus, we can decompose an \emph{off} period into slots in which
a secondary user succeeds and those in which no secondary user succeeds, i.e.,
$T_{\textrm{\emph{off}}} = T_s + T_{\textrm{\emph{ns}}}$.
Note that $T_{\textrm{\emph{on}}}$, $T_{\textrm{\emph{off}}}$, $T_{col}$, $T_s$, and $T_{\textrm{\emph{ns}}}$ are determined by the protocol
and the traffic arrival process whereas
$T_{pac}$ and $T_{int}$ are determined entirely by the traffic arrival process.

We explain how we (approximately) compute the performance metrics defined in this section.
The collision probability of the primary user can be computed as
$P_c = T_{col}/T_{\textrm{\emph{on}}}$ since the primary user transmits
whenever its state is \emph{on}. Also, the success
probability of the secondary users can be computed as $P_s = T_s/T_{\textrm{\emph{off}}}$.
The channel utilization of the primary user is given by $C_p = T_{pac}/T_{int}$,
while that of the secondary users is $C_s = T_s/T_{int}$.
The channel utilization of the system can be computed as $C = C_p + C_s = (T_{pac} + T_s)/T_{int}$.

When perfect control devices are available to broadcast the presence
of the primary user and to schedule access by the secondary users as in
\cite{chou}, we can obtain $T_{col} = 0$ and $T_s = T_{\textrm{\emph{off}}}$. Thus,
with control devices, we can achieve the maximum values of the performance
metrics $\overline{C}_p = T_{pac}/T_{int}$, $\overline{C}_s = (T_{int} - T_{pac})/T_{int}$,
and $\overline{C} = 1$.
Note that the channel utilization of the primary user is not affected by the absence of
control devices (as long as $T_{col} < T_{int} - T_{pac}$) although the primary user may experience increased delay
as $T_{col}$ becomes large due to contention between the primary user and the secondary users.
The value of $C_s$ becomes smaller as contention among the secondary users increases.
The ratio of $C$ to $\overline{C}$ can be used as a measure of inefficiency
due to the absence of control devices.

\subsection{Protocol Design Problem}

We formulate a problem solved by the protocol designer to determine
a protocol.
We assume that the protocol designer considers only non-intrusive protocols
with one-slot memory. Non-intrusiveness is a desirable property in that
it prevents the secondary users from interrupting the primary user once
the primary user obtains a successful transmission. We focus on protocols with one-slot
memory because they are simple to design and implement.
We also assume that the protocol designer has the most preferred fairness
level $\theta \in (0,1]$. Then non-intrusiveness together with fairness
level $\theta$ implies that $f(\textrm{\emph{success}}) = 1-\theta$ by \eqref{eq:fair}, and the remaining elements of
a protocol to be specified are transmission probabilities
following an idle state and a failure state, denoted by $q = f(\textrm{\emph{idle}})$ and
$r = f(\textrm{\emph{failure}})$, respectively. For simplicity, we call hereafter
a non-intrusive protocol with one-slot memory having fairness level $\theta$
a $\theta$-fair non-intrusive protocol.

The protocol designer aims to maximize the channel utilization
of the system while keeping the collision probability
of the primary user below a certain threshold level specified as $\eta \in (0,1)$.
The protection level $\eta$ can be considered as a requirement imposed by the primary user
or by spectrum regulators.
The protocol design problem can be formally expressed as
\begin{align*}
\max_{f \in \mathcal{F}} C \textrm{ subject to } P_c \leq \eta,
\end{align*}
where $\mathcal{F}$ is the set of all $\theta$-fair non-intrusive protocols.
Since $T_{pac}$ and $T_{int}$ are independent of the prescribed
protocol, the protocol design problem can be rewritten as
\begin{align} \label{eq:probprot}
\max_{(q,r) \in [0,1]^2} C_s = P_s \frac{T_{int} - T_{pac} - T_{col}}{T_{int}} \textrm{ subject to } T_{col} \leq \gamma,
\end{align}
where $\gamma = (\eta/(1-\eta)) T_{pac}$ is the threshold level for
$T_{col}$, derived from the relationship $P_c = T_{col}/(T_{pac}+T_{col})$
and the requirement $P_c \leq \eta$.
Note that $T_{col}$ appears both in the objective function and in the constraint.
The protocol designer prefers small $T_{col}$ for two reasons. Smaller
$T_{col}$ implies less interference to the primary user and at the same time
longer \emph{off} periods that the secondary users can utilize.
In Section IV we explain how to compute $P_s$ and $T_{col}$ analytically given
a $\theta$-fair non-intrusive protocol,
while in Section V we investigate the solution to the protocol design problem using numerical illustrations.

\section{Analytical Results}

\subsection{Derivation of the Success Probability of the Secondary Users}

We first study the operation of the system in an
\emph{off} period, in which the primary user is inactive.
To analyze performance in an \emph{off} period,
we construct a Markov chain whose state space is $\{0,1,\ldots,N\}$, where state $k$ represents
transmission outcomes in which exactly $k$ secondary users transmit.
The transition probability from state $k$ to state $k'$ in an \emph{off} period,
denoted $P_{\textrm{\emph{off}}}(k'|k)$, under a $\theta$-fair non-intrusive protocol is given by
\begin{eqnarray} \label{eq:off1}
P_{\textrm{\emph{off}}}(k'|0) &=& \binom{N}{k'} q^{k'} (1-q)^{N-k'} \quad \textrm{for $k'=0,\ldots,N$},\\
P_{\textrm{\emph{off}}}(k'|1) &=& \left\{ \begin{array}{ll}
\theta & \textrm{for $k'=0$}\\
1-\theta & \textrm{for $k'=1$}\\
0 & \textrm{for $k'=2,\ldots,N$},
\end{array} \right. \nonumber \\ 
P_{\textrm{\emph{off}}}(k'|k) &=& \left\{ \begin{array}{ll}
\binom{k}{k'} r^{k'} (1-r)^{k-k'}  & \textrm{for $k'=0,\ldots,k$}\\
0 & \textrm{for $k'=k+1,\ldots,N$}
\end{array} \right. ,\textrm{for $k=2,\ldots,N$}. \label{eq:off3}
\end{eqnarray}
The transition matrix of the Markov chain can be written in the form of
\begin{align*}
\mathbf{P}_{\textrm{\emph{off}}} = \kbordermatrix{&0&2& \cdots &N-1&N& & 1\\
0& * & * & \cdots & * & * & \vrule & *\\
2& * & * & \cdots & 0 & 0 & \vrule & *\\
\vdots & \vdots & \vdots & \ddots & \vdots & \vdots & \vrule & \vdots & \\
N-1& * & * & \cdots & * & 0 & \vrule & *\\
N& * & * & \cdots & * & * & \vrule & *\\
\cline{2-8}
1& \theta & 0 & \cdots & 0 & 0 & \vrule & 1-\theta
},
\end{align*}
where the entries marked with an asterisk can be found in \eqref{eq:off1} and \eqref{eq:off3}.

Consider a slot $t$ in which the state of the primary user has changed from \emph{on} to \emph{off},
i.e., $y_p^{t-1} = on$ and $y_p^t = \textrm{\emph{off}}$, where $y_p^{\tau}$ is the state of the primary
user in slot $\tau$.
Since such a transition can occur
only if the primary user transmitted a packet successfully in slot $t-1$, it must be the case that
$y_i^{t-1} = busy$ for every secondary user $i$,
where $y_i^{\tau}$ is the state of secondary user $i$ in slot $\tau$.
By non-intrusiveness, no secondary user transmits in slot $t$, and thus an \emph{off} period
always begins with an idle slot (state 0).
Starting from an idle slot, the secondary users contend with each other
until a secondary user obtains a success, i.e., state 1 is reached. When a secondary
user obtains a success, it transmits with probability $1-\theta$ in the next slot
while all the other secondary users wait. A period of consecutive successes by a secondary user
ends with an idle slot, when the successful user waits. In short, an \emph{off} period can be considered as
the alternation of a contention period and a success period, which is continued until
traffic arrives to the primary user. A success period consists of
slots with consecutive successes by a secondary user, whereas a contention period
begins with an idle slot and lasts until a secondary user succeeds. Since all the secondary users transmit
with the same transmission probability following an idle slot, they have an equal
chance of becoming a successful user for the following success period at the point when a contention period starts.

Let $\tilde{T}_s$ and $\tilde{T}_{\textrm{\emph{ns}}}$ be the average duration
(measured in slots) of a success period and a contention period, respectively.
$\tilde{T}_s$ is determined by the fairness level $\theta$, where the relationship is given by
$\tilde{T}_s = 1/\theta$.
Let $\mathbf{Q}_{\textrm{\emph{off}}}$ be the $N$-by-$N$ matrix in the upper-left
corner of $\mathbf{P}_{\textrm{\emph{off}}}$. Suppose that $0 < q,r < 1$ so that all the
entries of $\mathbf{P}_{\textrm{\emph{off}}}$ marked with an asterisk are nonzero.
Then $(\mathbf{I} - \mathbf{Q}_{\textrm{\emph{off}}})^{-1}$ exists and is called the
fundamental matrix for $\mathbf{P}_{\textrm{\emph{off}}}$,
when state 1 is absorbing (i.e., $\theta = 0$) \cite{grinstead}.
The average number of slots in state $k \neq 1$ starting from state 0 (an idle slot) is given by
the $(1,k)$-entry of $(\mathbf{I} - \mathbf{Q}_{\textrm{\emph{off}}})^{-1}$. Hence,
the average number of slots to hit state 1 (a success slot) for the first time
starting from an idle slot is given by
the first entry of $(\mathbf{I} - \mathbf{Q}_{\textrm{\emph{off}}})^{-1} \mathbf{e}$,
where $\mathbf{e}$ is a column vector of length $N$ all of whose entries are 1.
Hence, we obtain $\tilde{T}_{\textrm{\emph{ns}}} = [(\mathbf{I} - \mathbf{Q}_{\textrm{\emph{off}}})^{-1} \mathbf{e}]_1$,
where $[\mathbf{v}]_k$ denotes the $k$-th entry of vector $\mathbf{v}$.
Note that $\tilde{T}_{\textrm{\emph{ns}}}$ is independent of $\theta$. That is, the average duration of a contention
period is not affected by the average duration of a success period.
The success probability of the secondary users can be computed by
\begin{align} \label{eq:psuc}
P_s = \frac{\tilde{T}_s}{\tilde{T}_{\textrm{\emph{ns}}} + \tilde{T}_s}
= \frac{1}{\theta [(\mathbf{I} - \mathbf{Q}_{\textrm{\emph{off}}})^{-1} \mathbf{e}]_1 + 1},
\end{align}
for $(q,r) \in (0,1)^2$.

An alternative method to compute the success probability of the secondary users
is to use a stationary distribution. Since $\theta \in (0,1]$, all states
communicate with each other under the transition matrix $\mathbf{P}_{\textrm{\emph{off}}}$
for all $(q,r) \in (0,1)^2$.
Hence, the Markov chain is irreducible, and there exists
a unique stationary distribution $\mathbf{w}_{\textrm{\emph{off}}}$,
which satisfies
\begin{align} \label{eq:wnorm}
\mathbf{w}_{\textrm{\emph{off}}} = \mathbf{w}_{\textrm{\emph{off}}} \mathbf{P}_{\textrm{\emph{off}}}& \textrm{ and }
\mathbf{w}_{\textrm{\emph{off}}} \mathbf{e} = 1.
\end{align}
Let ${w}_{\textrm{\emph{off}}}(k)$ be the entry of $\mathbf{w}_{\textrm{\emph{off}}}$
corresponding to state $k$, for $k = 0,1,\ldots,N$.
Then ${w}_{\textrm{\emph{off}}}(k)$ gives the probability of state $k$ during an \emph{off} period.
In particular, the success probability of the secondary users
is given by ${w}_{\textrm{\emph{off}}}(1)$.
Since contention and success periods alternate from the beginning
of an \emph{off} period, the stationary distribution yields the probabilities of states
for any duration of an \emph{off} period (assuming that $T_{\textrm{\emph{off}}}$ is sufficiently larger than
$\tilde{T}_{\textrm{\emph{ns}}} + \tilde{T}_s$), not just the limiting probabilities as an \emph{off} period
lasts infinitely long. By manipulating \eqref{eq:wnorm}, we can derive that
${w}_{\textrm{\emph{off}}}(1) = P_s$, whose expression is given in \eqref{eq:psuc}.

\subsection{Derivation of the Collision Probability of the Primary User}

We next study the operation of the system in an
\emph{on} period, in which the primary user always transmits.
To analyze performance in an \emph{on} period,
we construct another Markov chain with the same state space $\{0,1,\ldots,N\}$
as before. Again, state $k$ corresponds to transmission outcomes in which exactly $k$ secondary users transmit.
The transition probability from state $k$ to state $k'$ in an \emph{on} period, denoted
$P_{\textrm{\emph{on}}}(k'|k)$, under a $\theta$-fair non-intrusive protocol is given by
\begin{eqnarray} \label{eq:on}
P_{\textrm{\emph{on}}}(k'|k) &=& \left\{ \begin{array}{ll}
\binom{k}{k'} r^{k'} (1-r)^{k-k'} & \textrm{for $k'=0,\ldots,k$}\\
0 & \textrm{for $k'=k+1,\ldots,N$}
\end{array} \right. ,\textrm{for $k=0,\ldots,N$}.
\end{eqnarray}
The transition matrix of the Markov chain can be written in the form of
\begin{align*}
\mathbf{P}_{\textrm{\emph{on}}} = \kbordermatrix{&1&2& \cdots &N-1&N& & 0\\
1& * & 0 & \cdots & 0 & 0 & \vrule & *\\
2& * & * & \cdots & 0 & 0 & \vrule & *\\
\vdots & \vdots & \vdots & \ddots & \vdots & \vdots & \vrule & \vdots & \\
N-1& * & * & \cdots & * & 0 & \vrule & *\\
N& * & * & \cdots & * & * & \vrule & *\\
\cline{2-8}
0& 0 & 0 & \cdots & 0 & 0 & \vrule & 1
},
\end{align*}
where the entries marked with an asterisk can be found in \eqref{eq:on}.
Note that state 0, which corresponds to a success by the primary user,
is absorbing because once the primary user has a successful transmission,
its transmissions in the following slots are not interrupted by the secondary users.
Hence, collisions in an \emph{on} period occur only before the primary
user obtains a successful transmission. Also,
the average number of collisions experienced by the primary
user in an \emph{on} period, $T_{col}$,
is independent of the length of traffic, $T_{pac}$.
Let $\mathbf{Q}_{\textrm{\emph{on}}}$ be the $N$-by-$N$ matrix in the upper-left corner of $\mathbf{P}_{\textrm{\emph{on}}}$.
For $r \in [0,1)$, the matrix $\mathbf{I} - \mathbf{Q}_{\textrm{\emph{on}}}$ is invertible, and
the average number of slots until the first success by the primary user starting from
state $k$ is given by the $k$-th entry of
$(\mathbf{I} - \mathbf{Q}_{\textrm{\emph{on}}})^{-1} \mathbf{e}$, for $k = 1, \ldots, N$.

Consider a slot $t$ in which the state of the primary user has changed from \emph{off} to \emph{on},
i.e., $y_p^{t-1} = \textrm{\emph{off}}$ and $y_p^t = on$.
Then an \emph{on} period begins from slot $t$.
The number of collisions that the primary user
expect to experience in the \emph{on} period depends on the transmission outcome in slot $t-1$,
the last slot of the preceding \emph{off} period.
Suppose that there was a collision among $k \geq 2$ secondary users in slot $t-1$.
Then the Markov chain starts from state $k$ in slot $t-1$.
Since the \emph{on} period starts in slot $t$, the number of collisions in the \emph{on} period does not include
the collision in slot $t-1$. Hence, the average number of collisions
until the first success in an \emph{on} period when the preceding \emph{off} period
ended with $k$ transmissions is given by
\begin{align*}
d(k) = [ (\mathbf{I} - \mathbf{Q}_{\textrm{\emph{on}}})^{-1} \mathbf{e} ]_k - 1,
\end{align*}
for $k = 2, \ldots, N$.

Suppose that there was a success in slot $t-1$. Then the successful secondary user transmits
with probability $1-\theta$ while all the other secondary users wait in slot $t$.
Thus, with probability $\theta$, the primary user succeeds in slot $t$,
and with probability $1-\theta$, state 1 occurs in slot $t$, from which
it takes $[ (\mathbf{I} - \mathbf{Q}_{\textrm{\emph{on}}})^{-1} \mathbf{e} ]_1$ collisions
on average to reach a success by the primary user. Therefore,
the average number of collisions
until the first success in an \emph{on} period when the preceding \emph{off} period
ended with a success is given by
\begin{align} \label{eq:d1}
d(1) = \theta \cdot 0 + (1-\theta)  [ (\mathbf{I} - \mathbf{Q}_{\textrm{\emph{on}}})^{-1} \mathbf{e} ]_1
= (1-\theta)  [ (\mathbf{I} - \mathbf{Q}_{\textrm{\emph{on}}})^{-1} \mathbf{e} ]_1.
\end{align}

Suppose that slot $t-1$ was idle. Then
with probability $\binom{N}{k} q^k (1-q)^{N-k}$, slot $t$ contains transmission by $k$ secondary users, for $k = 0, \ldots, N$.
With probability $(1-q)^N$ the primary user experiences no collision while
with probability $\binom{N}{k} q^k (1-q)^{N-k}$ the \emph{on} period begins with state $k$,
for $k = 1,\ldots, N$. Therefore,
the expected number of collisions
until the first success in an \emph{on} period when the preceding \emph{off} period
ended with an idle slot is given by
\begin{align*}
d(0) &= (1-q)^N \cdot 0 + \sum_{k=1}^N \binom{N}{k} q^k (1-q)^{N-k} [ (\mathbf{I} - \mathbf{Q}_{\textrm{\emph{on}}})^{-1} \mathbf{e} ]_k\\
&= \sum_{k=1}^N \binom{N}{k} q^k (1-q)^{N-k} [ (\mathbf{I} - \mathbf{Q}_{\textrm{\emph{on}}})^{-1} \mathbf{e} ]_k.
\end{align*}

\begin{figure}[hb]
\begin{center}
\includegraphics[width=0.8\textwidth]{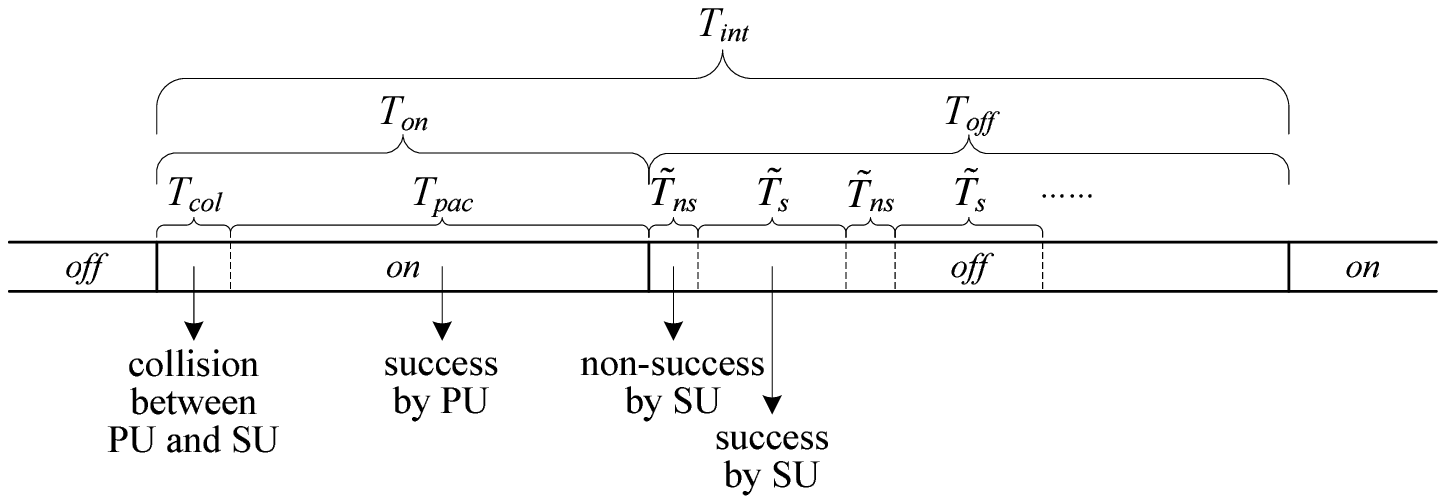}
\caption{Operation of the system under a $\theta$-fair non-intrusive protocol.}
\label{fig:oper}
\end{center}
\end{figure}

The probability that the last slot of an \emph{off} period
has $k$ transmissions is given by ${w}_{\textrm{\emph{off}}}(k)$, for $k = 0,1,\ldots,N$.
Hence, the average number of collisions that the primary user experiences
before its first success in an \emph{on} period is given by
\begin{align*}
T_{col} = \sum_{k=0}^{N} {w}_{\textrm{\emph{off}}}(k) d(k).
\end{align*}
Once the primary user succeeds in an \emph{on} period, it
has successful transmissions until it finishes transmitting
all the packets it has, from which point an \emph{off} period begins.
Using the relationship $P_c = T_{col}/(T_{pac}+T_{col})$, we can
compute the collision probability of the primary user.
The operation of the system under a $\theta$-fair non-intrusive protocol is summarized in Fig.~\ref{fig:oper}.

\section{Numerical Results}

\subsection{Graphical Illustration of the Protocol Design Problem} \label{sec:graph}

Based on the results in Section IV, we can show that, for a given fairness level $\theta \in (0,1]$,
$P_s$ and $T_{col}$ are continuous functions of $(q,r)$ on the
interior of $[0,1]^2$. In order to guarantee the existence of a solution,
in this section we consider the protocol design problem on a restricted domain,
\begin{align} \label{eq:pdprob2}
\max_{(q,r) \in [\epsilon,1-\epsilon]^2} C_s = P_s \frac{T_{int} - T_{pac} - T_{col}}{T_{int}} \textrm{ subject to } T_{col} \leq \gamma,
\end{align}
for a small $\epsilon > 0$. Throughout this section, we set $\epsilon = 10^{-4}$.
We say that a protocol is optimal if it solves \eqref{eq:pdprob2}.
An optimal protocol gives an approximate, if not exact, solution to \eqref{eq:probprot}.

\begin{figure}%
\centering
\subfloat[$P_s$]{%
\label{fig:cont-a}%
\includegraphics[width=0.45\textwidth]{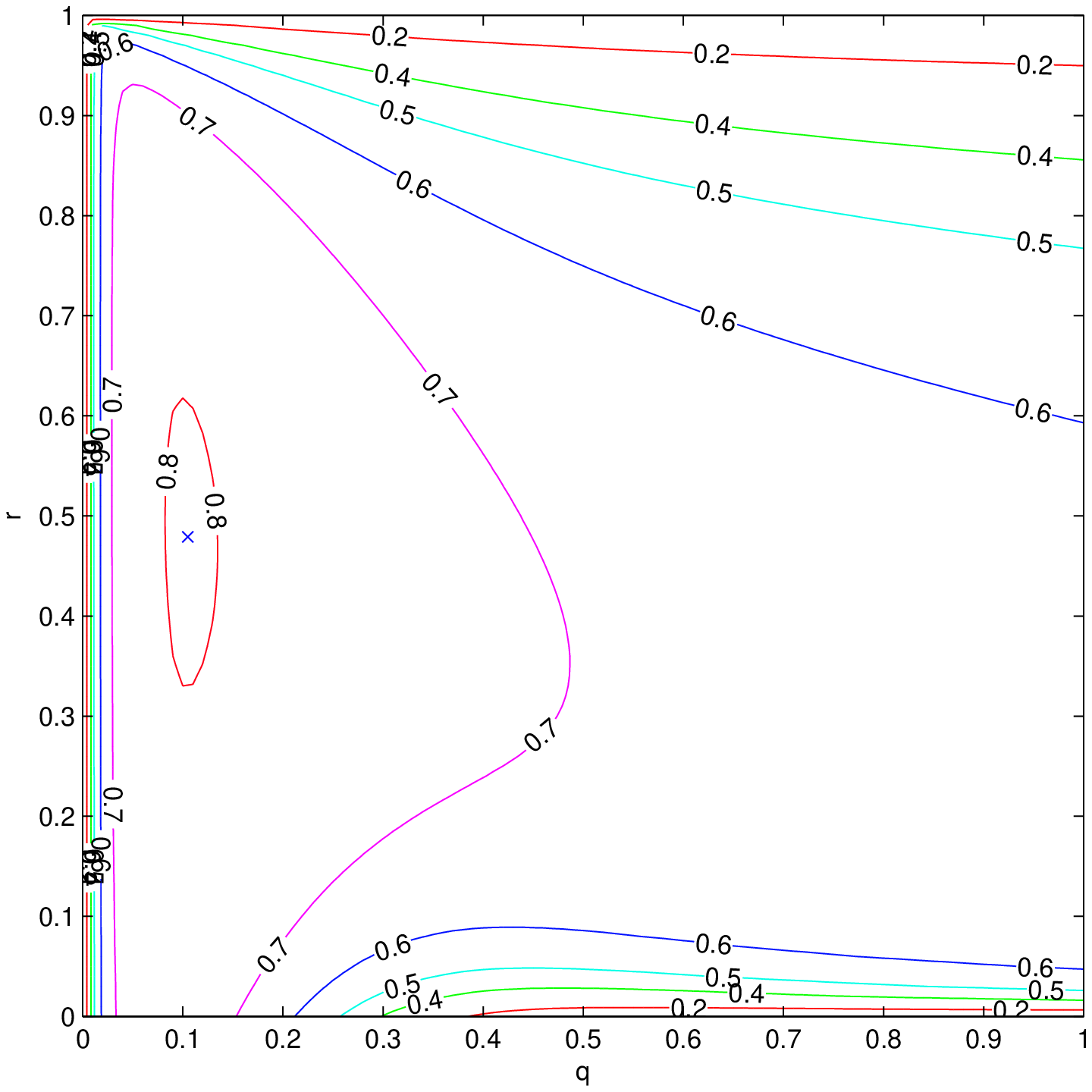}}%
\subfloat[$T_{col}$]{%
\label{fig:cont-b}%
\includegraphics[width=0.45\textwidth]{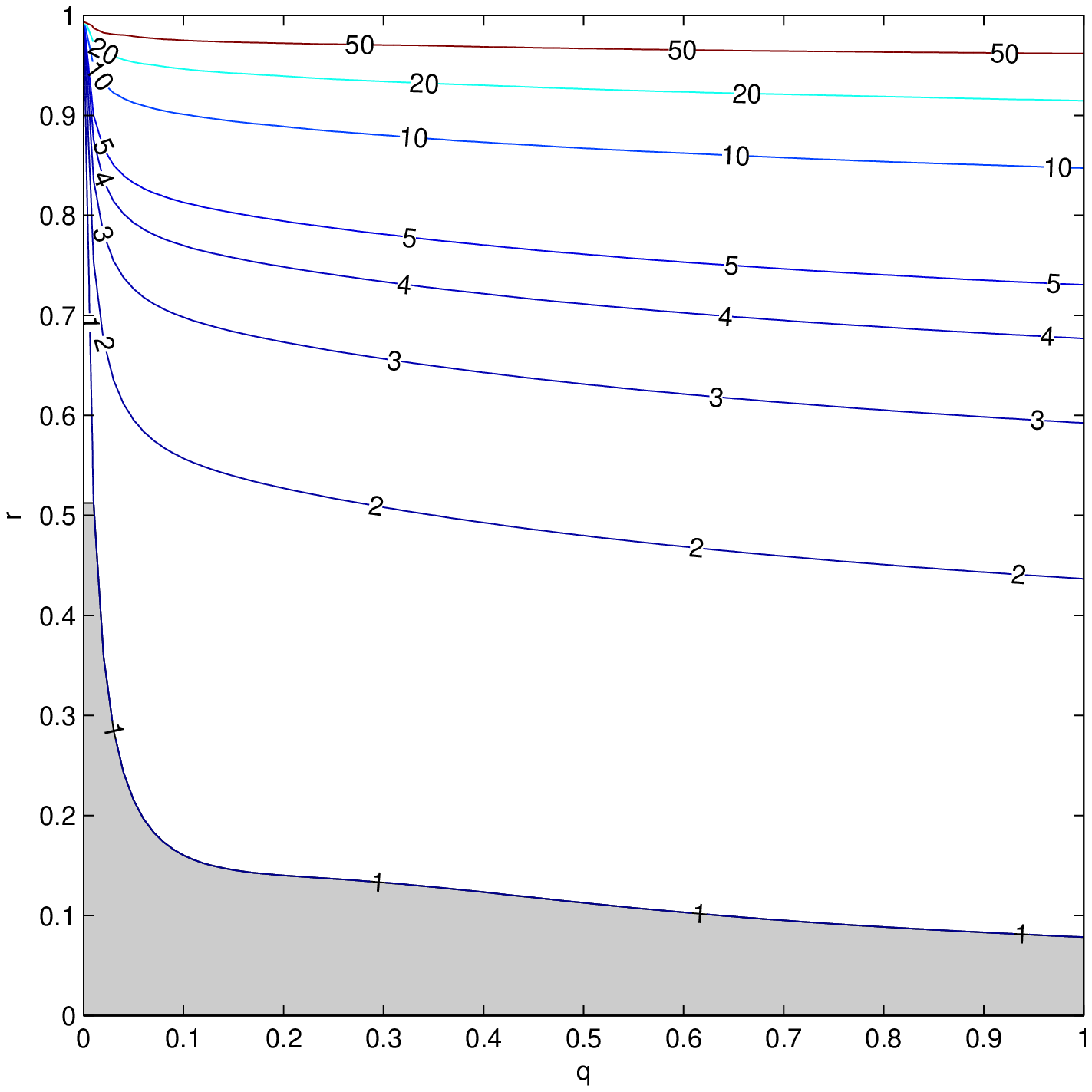}}\\
\subfloat[$C_s$]{%
\label{fig:cont-c}%
\includegraphics[width=0.45\textwidth]{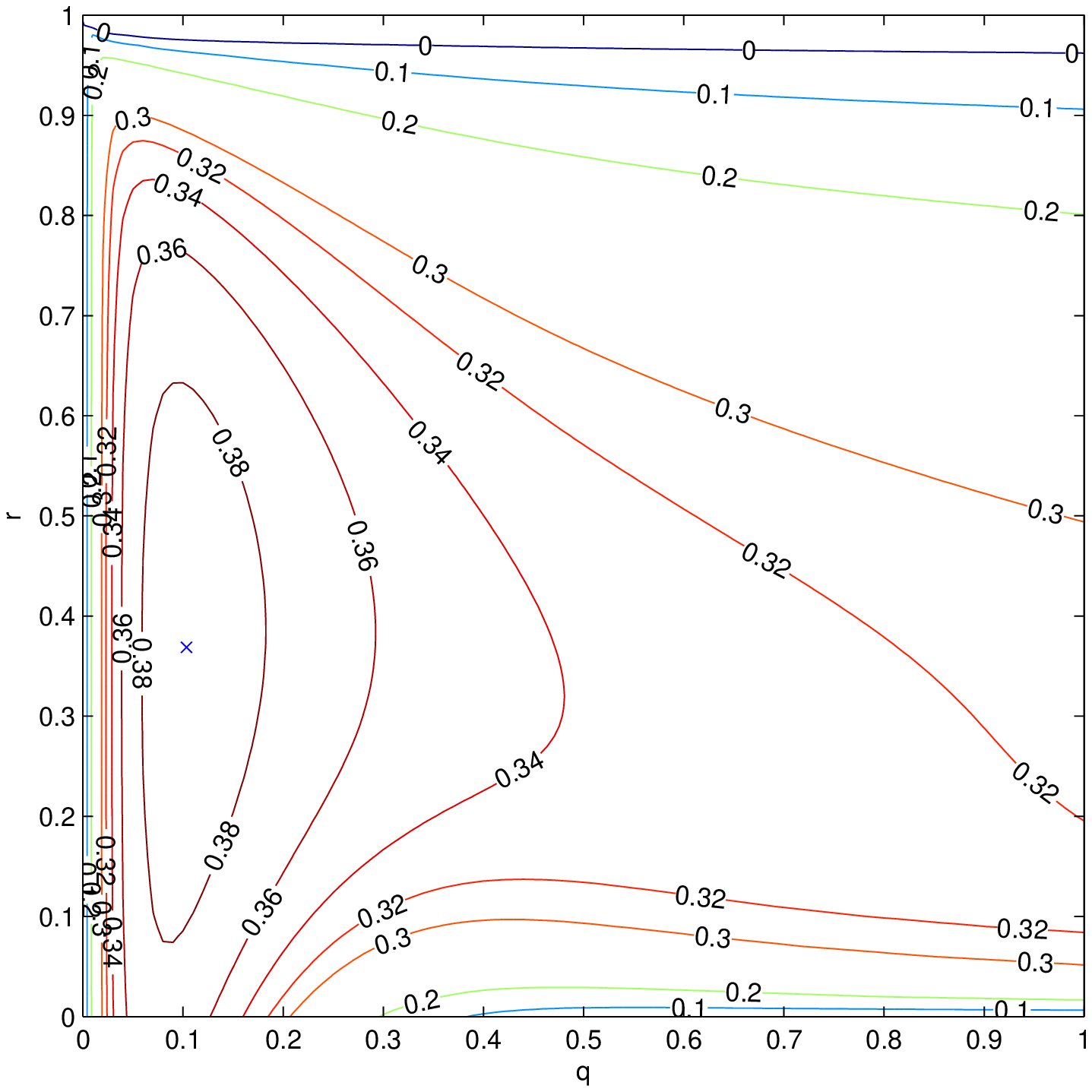}}%
\caption{Contour curves of
$P_s$, $T_{col}$, and $C_s$ as functions of $(q,r)$
when $N = 10$, $T_{int} = 100$, $T_{pac} = 50$, and $\theta = 0.1$.}
\label{fig:contour}%
\end{figure}

In Fig.~\ref{fig:contour}, we show the dependence of the performance metrics,
$P_s$, $T_{col}$, and $C_s$, on the protocol $(q,r)$. To obtain the results, we consider a network with
$N = 10$, $T_{int} = 100$, and $T_{pac} = 50$, and set $\theta = 0.1$. The maximum value of
$C_s$ is thus $0.5$, while $\tilde{T}_s = 10$.
Fig.~\ref{fig:contour}\subref{fig:cont-a} plots
the contour curves of $P_s$. The success probability of the secondary users $P_s$
is maximized at $q = 0.11$ and $r = 0.48$, and the maximum value
of $P_s$ is $0.804$, which corresponds to the minimum value of $\tilde{T}_{\textrm{\emph{ns}}}$
as $2.44$.
The value of $(q,r)$ that maximizes $P_s$ can be justified as follows. Following an idle slot
in an \emph{off} period, every secondary user transmits with probability $q$, and thus the probability of success
is maximized when $q = 1/N$ \cite{massey}. During an \emph{off} period, a collision cannot follow a success, and
following an idle slot, a collision involving two transmissions is most likely among all kinds of collisions
when $q \approx 1/N$. Since non-colliding users do not transmit following a collision under a non-intrusive protocol,
the probability of success between two contending users is maximized when $r = 1/2$.
$r$ is chosen slightly smaller than $1/2$ because collisions involving
more than two transmissions occur with small probability.

Fig.~\ref{fig:contour}\subref{fig:cont-b} plots the contour curves of $T_{col}$.
As $q$ and $r$ are large, secondary users transmit aggressively in a contention period, intensifying
interference to the primary user when it starts transmitting. Thus, $T_{col}$ is increasing in both $q$ and $r$.
The set of $(q,r)$ that satisfies the constraint $T_{col} \leq \gamma$ can be represented by the region below the
contour curve of $T_{col}$ at level $\gamma$. For example, the shaded area
in Fig.~\ref{fig:contour}\subref{fig:cont-b} represents the constraint set corresponding to $T_{col} \leq 1$.
Since $P_c = T_{col}/(T_{pac}+T_{col})$, $P_c$ is monotonically increasing
in $T_{col}$, and thus the contour curves of $P_c$ have the same shape
as those of $T_{col}$.

Fig.~\ref{fig:contour}\subref{fig:cont-c} plots the contour curves of $C_s$.
Let $(q^*, r^*) = \arg \max_{(q,r) \in [\epsilon,1-\epsilon]^2} C_s$. That is, $(q^*, r^*)$ represents
the $\theta$-fair non-intrusive protocol that maximizes the channel utilization of the secondary users when no constraint
is imposed on the collision probability of the primary user.
Note that the channel utilization of the secondary users can be expressed as
$C_s = P_s \times P_{\textrm{\emph{off}}}$, where $P_{\textrm{\emph{off}}}$ is the proportion of ``\emph{off}'' slots, i.e.,
$P_{\textrm{\emph{off}}} = T_{\textrm{\emph{off}}}/T_{int} = (T_{int} - T_{pac} - T_{col})/T_{int}$.
Hence, in order to maximize $C_s$, we need to take into account both $P_s$ and $P_{\textrm{\emph{off}}}$.
To maximize $P_s$, $(q,r)$ needs to be chosen at $(0.11, 0.48)$. Since
$P_{\textrm{\emph{off}}}$ is decreasing in $T_{col}$, maximizing $P_{\textrm{\emph{off}}}$
requires $(q,r)$ to be $(\epsilon,\epsilon)$, at which $T_{col}$ is minimized. In Fig.~\ref{fig:contour}\subref{fig:cont-c},
it is shown that this conflict is resolved by choosing $(q,r)$ somewhere in between.
The protocol that maximizes the channel utilization of the secondary users is given by
$(q^*,r^*) = (0.10, 0.37)$, while the maximum value of $C_s$ is $0.390$.

\begin{figure}
\begin{center}
\includegraphics[width=0.7\textwidth]{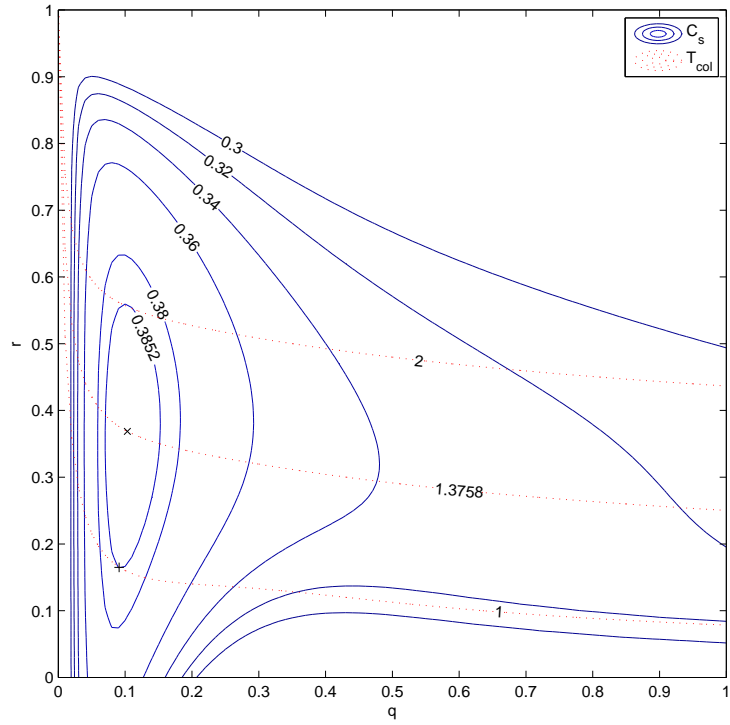}
\caption{Illustration of optimal protocols.}
\label{fig:contmap}
\end{center}
\end{figure}

Fig.~\ref{fig:contmap} shows the contour curves of $C_s$ and $T_{col}$ in
the same graph to illustrate the protocol design problem \eqref{eq:pdprob2}.
The protocol design problem is to find
the largest value of $C_s$ on the region of $(q,r)$ that satisfies $T_{col} \leq \gamma$.
Let $\gamma^*$ be the value of $T_{col}$ at $(q^*, r^*)$.
With the parameter specification to obtain Fig.~\ref{fig:contmap}, we have $\gamma^* = 1.376$.
We say that a constraint is binding if its removal results in a strict improvement
in the objective value and non-binding otherwise. Then the constraint in \eqref{eq:pdprob2}
is binding if $\gamma < \gamma^*$ and non-binding if $\gamma \geq \gamma^*$.
For example, if $\gamma = 1$, the constraint is binding and the optimal protocol
is given by the point on the contour curve of $T_{col}$ at level $1$, marked
with `$+$' in Fig.~\ref{fig:contmap}, where
a contour curve of $T_{col}$ and that of $C_s$ are tangent to each other.
In contrast, if $\gamma = 2$, the constraint is non-binding and
the optimal protocol is given by the solution to the unconstrained problem, $(q^*,r^*) = (0.10, 0.37)$, marked
with `$\times$' in Fig.~\ref{fig:contmap}.

\begin{figure}%
\centering
\subfloat[][]{%
\label{fig:gvary-a}%
\includegraphics[width=0.5\textwidth]{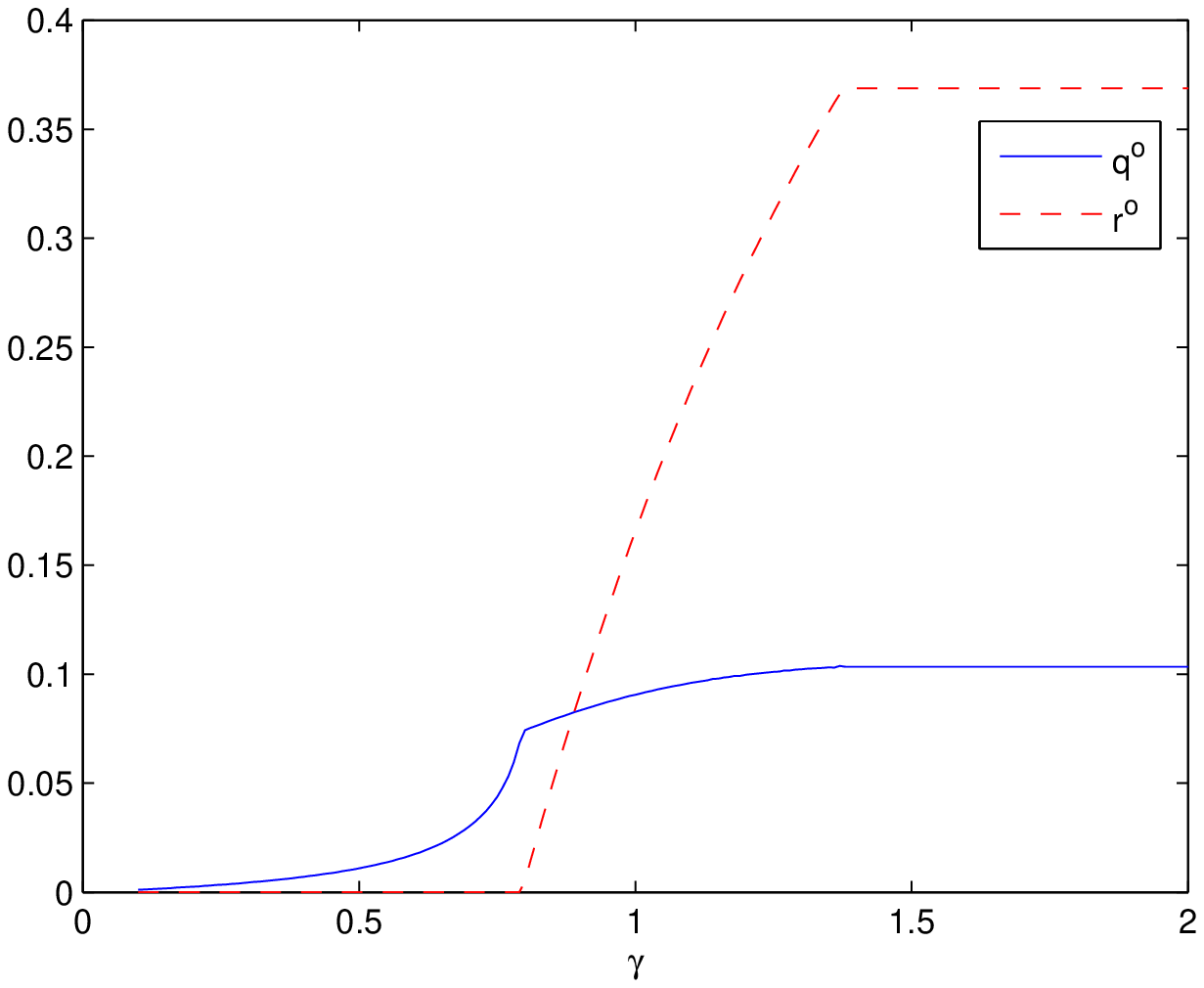}}%
\subfloat[][]{%
\label{fig:gvary-b}%
\includegraphics[width=0.5\textwidth]{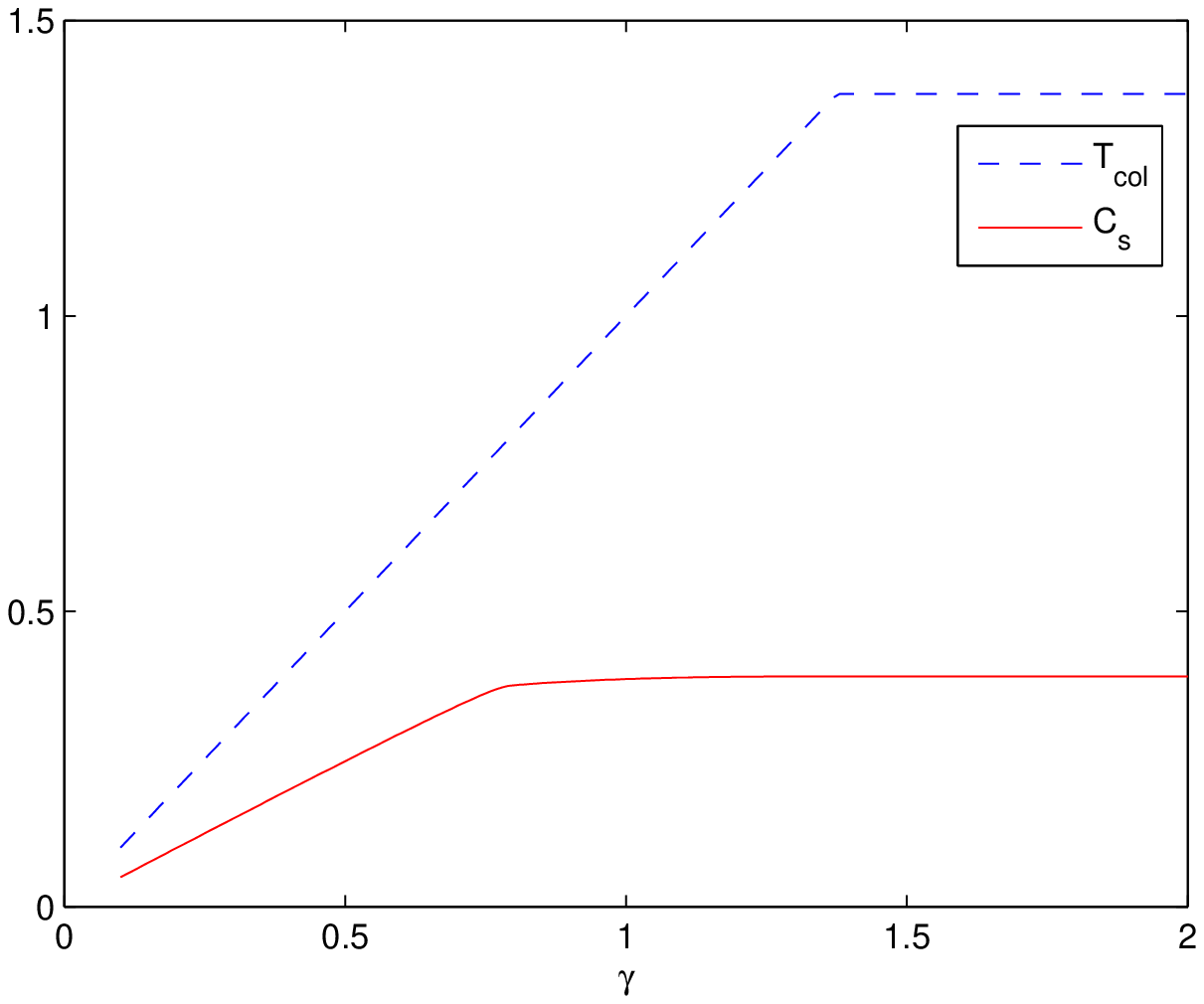}}%
\caption{Solution to the protocol design problem
for $\gamma$ between $0.1$ and $2$ when $N = 10$, $T_{int} = 100$, $T_{pac} = 50$, and $\theta = 0.1$:
\protect\subref{fig:gvary-a} optimal protocols, and
\protect\subref{fig:gvary-b} the values of $T_{col}$ and $C_s$ at the optimal protocols.}
\label{fig:gvary}%
\end{figure}

Fig.~\ref{fig:gvary} shows the solutions to the protocol design problem
for $\gamma$ between $0.1$ and $2$. Fig.~\ref{fig:gvary}\subref{fig:gvary-a}
plots optimal protocols, denoted by $(q^o,r^o)$, as $\gamma$ varies while
Fig.~\ref{fig:gvary}\subref{fig:gvary-b} shows the values of $T_{col}$ and $C_s$
at the optimal protocols. We can divide the range of $\gamma$ into three regions:
$(0, 0.8]$, $(0.8, 1.38)$, and $[1.38, \infty)$.
For $\gamma \leq 0.8$, the optimal protocol occurs at the corner with $r^o = \epsilon$. As $\gamma$ decreases
in this region, $q^o$ decreases to $\epsilon$ while $r^o$ stays at $\epsilon$, which makes $C_s$ decrease to 0.
Smaller $\gamma$ means that transmissions by the primary user are less interfered, and this can be achieved
by inhibiting transmissions by the secondary users. For $\gamma \in (0.8, 1.38)$, the solution to
the protocol design problem is interior while the constraint $T_{col} \leq \gamma$ is still
binding. The trade-off between $T_{col}$ and $C_s$ is less severe in this region
than in $(0, 0.8]$. Reducing $\gamma$ from $1.38$ to $0.8$ results in a slight decrease in $C_s$
from $0.39$ to $0.37$. For $\gamma \geq 1.38$, the constraint
$T_{col} \leq \gamma$ is non-binding, and thus $(q^o,r^o)$ remains at $(q^*,r^*) = (0.10, 0.37)$
while $C_s$ remains at its unconstrained maximum level, $0.39$.
The rate of change in the maximum value of $C_s$ with respect to $\gamma$ suggests that
keeping $T_{col}$ below $0.8$ induces a large cost in terms of the reduced channel utilization,
maintaining $T_{col}$ between $0.8$ and $1.38$ only a minor cost,
and tolerating $T_{col}$ larger than $1.38$ no cost.
In other words, when the optimal solution to the protocol design
problem is interior, the optimal dual variable on the constraint $T_{col} \leq \gamma$ is close
to zero or is zero.

\subsection{Varying the Number of Secondary Users}

\begin{figure}%
\centering
\subfloat[][]{%
\label{fig:nunc-a}%
\includegraphics[width=0.5\textwidth]{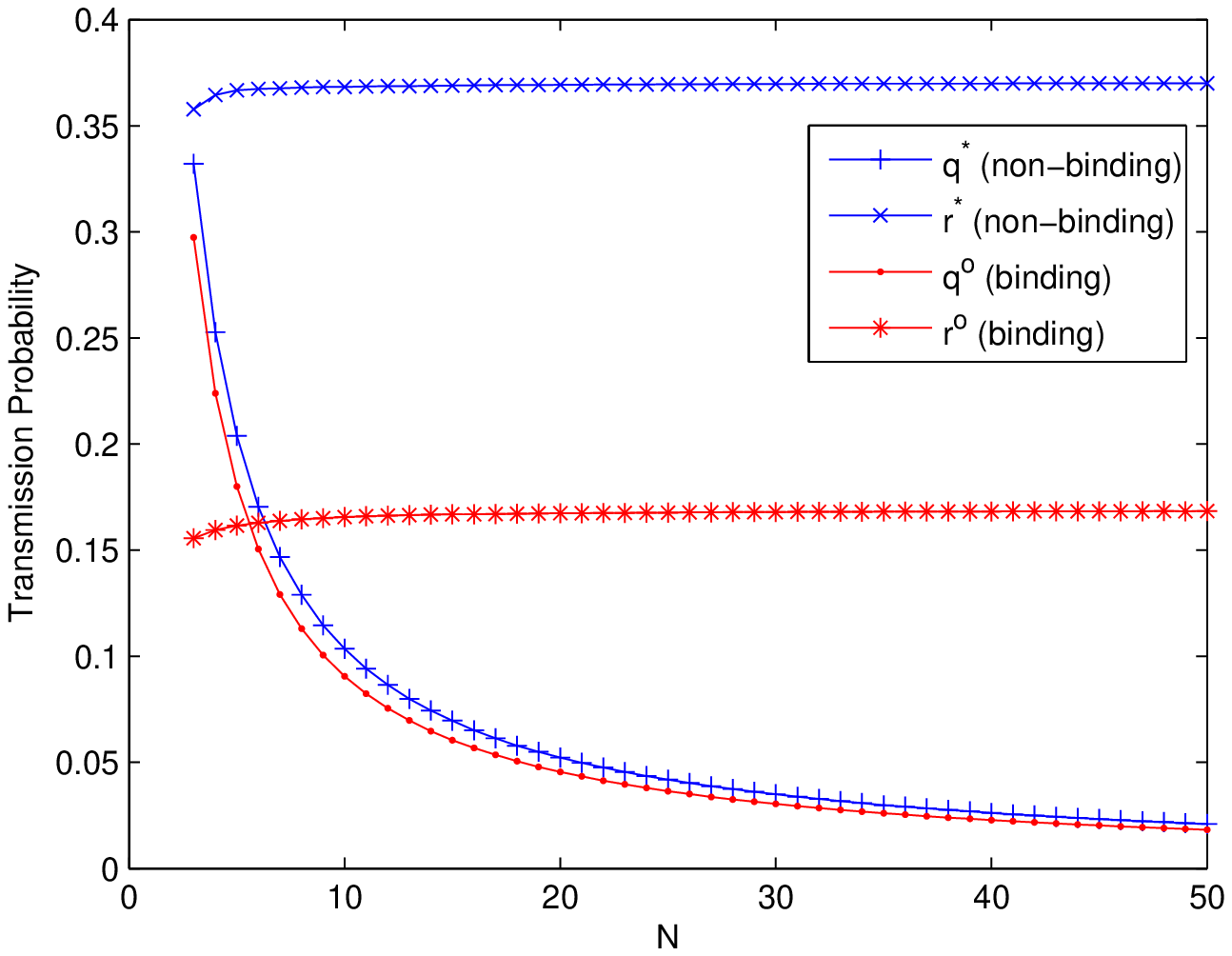}}%
\subfloat[][]{%
\label{fig:nunc-b}%
\includegraphics[width=0.5\textwidth]{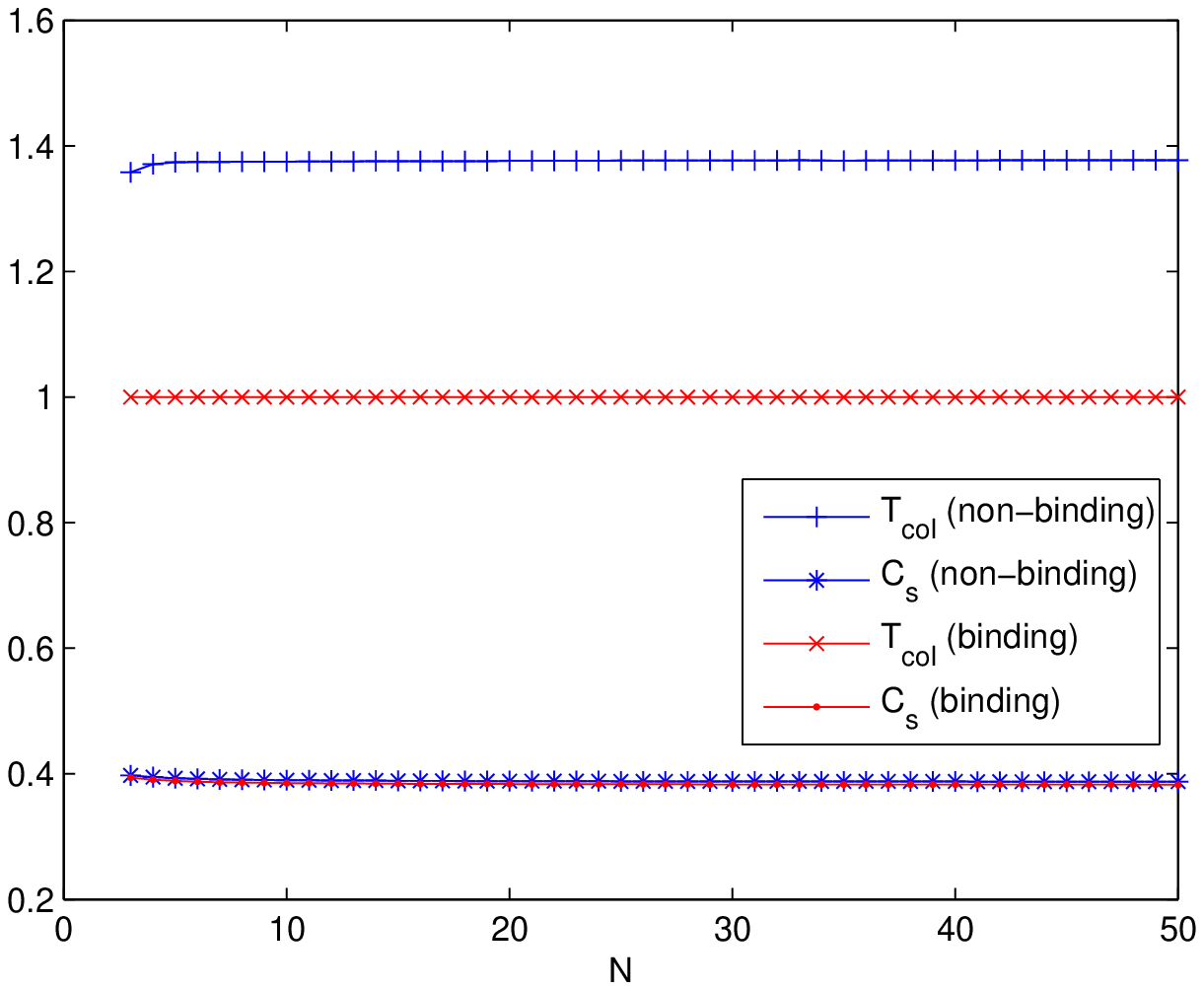}}%
\caption{Solution to the protocol design problem
for $N$ between $3$ and $50$ when $T_{int} = 100$, $T_{pac} = 50$, and $\theta = 0.1$:
\protect\subref{fig:nunc-a} optimal protocols, and
\protect\subref{fig:nunc-b} the values of $T_{col}$ and $C_s$ at the optimal protocols.}
\label{fig:nunc}%
\end{figure}

We study how the solution to the protocol design problem changes as the number of
secondary users varies between $3$ and $50$. We fix other parameters of the model as before. We first solve the
protocol design problem with a non-binding constraint, assuming that $\gamma$ is sufficiently large.
Fig.~\ref{fig:nunc}\subref{fig:nunc-a} shows optimal
protocols $(q^*,r^*)$ when the constraint is non-binding. As $N$ increases from $3$ to $50$, $q^*$ decreases
from $0.33$ to $0.02$ while $r^*$ increases from $0.36$ to $0.37$.
Fig.~\ref{fig:nunc}\subref{fig:nunc-b} plots the values of $T_{col}$ and $C_s$
at $(q^*,r^*)$. As $N$ increases from $3$ to $50$,
$T_{col}$ increases from $1.36$ to $1.38$ while
$C_s$ decreases from $0.40$ to $0.39$.
The results show that when the constraint is non-binding,
the degree of contention increases with the number of the secondary users
but only slightly, as the values of $T_{col}$ and $C_s$ are almost constant as $N$ varies.
Almost constant $T_{col}$ implies that, even without a constraint on $T_{col}$, interruption to the primary user
can be kept below a certain level.
This is because under optimal protocols the primary user is likely to contend with
at most two secondary users when it starts transmitting, regardless of the total number
of the secondary users.
Also, almost constant $C_s$ implies that optimal protocols are capable of resolving contention among
the secondary users efficiently even if there are many secondary users sharing the channel.
The values of $T_{col}$ at $(q^*,r^*)$
can be interpreted as the minimum values of $\gamma$ that make the constraint of
the protocol design problem non-binding.

Now we set $\gamma = 1$ so that the constraint is binding for all $N$ between $3$ and $50$.
Fig.~\ref{fig:nunc}\subref{fig:nunc-a} shows optimal
protocols $(q^o,r^o)$ when the constraint is given by $T_{col} \leq 1$. As $N$ increases from $3$ to $50$, $q^o$ decreases
from $0.30$ to $0.02$ while $r^o$ increases from $0.16$ to $0.17$.
Imposing the constraint limits the values of $q$ and $r$,
but it impacts $r$ more than $q$, i.e., $q^o \approx q^*$ and $r^o < r^*$ for given $N$,
due to the shape of the contour curves of $C_s$ as illustrated in Fig.~\ref{fig:contmap}.
Fig.~\ref{fig:nunc}\subref{fig:nunc-b} plots the values of $T_{col}$ and $C_s$
at $(q^o,r^o)$. As $N$ increases from $3$ to $50$,
$T_{col}$ stays at $1$, confirming that the constraint $T_{col} \leq 1$ is binding,
while $C_s$ decreases from $0.39$ to $0.38$.
Again, $C_s$ is almost constant with respect to $N$ even when a constraint is imposed on $T_{col}$.
We can see that requiring $T_{col} \leq 1$ decreases the maximum values of $C_s$
only slightly because the constraint with $\gamma = 1$ is mild so that
the optimal protocols remain interior. If we impose a sufficiently strong constraint,
i.e., choose a small $\gamma$, then we have the optimal protocol at the corner, $q^o < q^*$ and $r^o = \epsilon$,
and $C_s$ is reduced significantly, as suggested in Fig.~\ref{fig:gvary}.

\subsection{Varying the Fairness Level}

\begin{figure}%
\centering
\subfloat[][]{%
\label{fig:tunc-a}%
\includegraphics[width=0.5\textwidth]{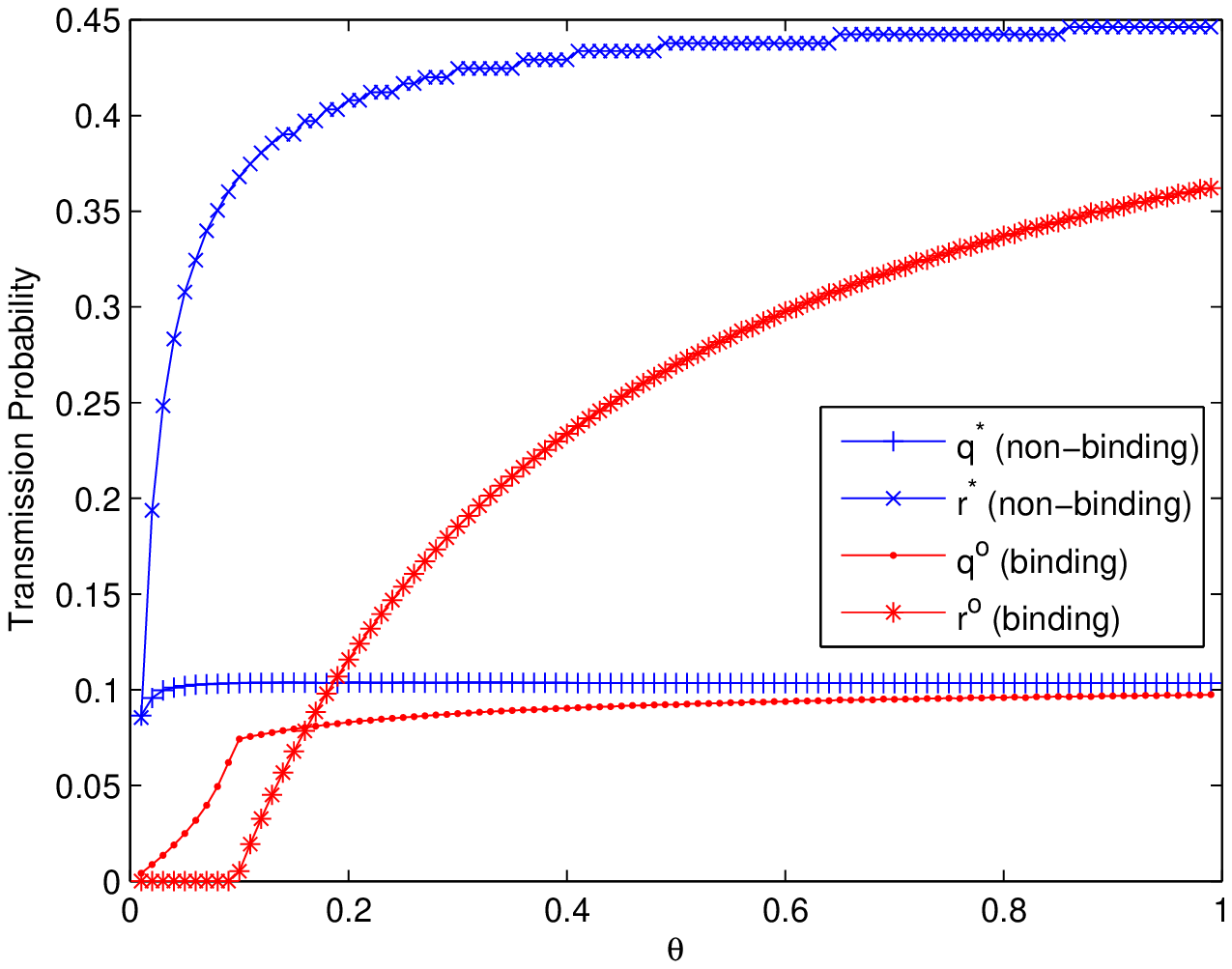}}%
\subfloat[][]{%
\label{fig:tunc-b}%
\includegraphics[width=0.5\textwidth]{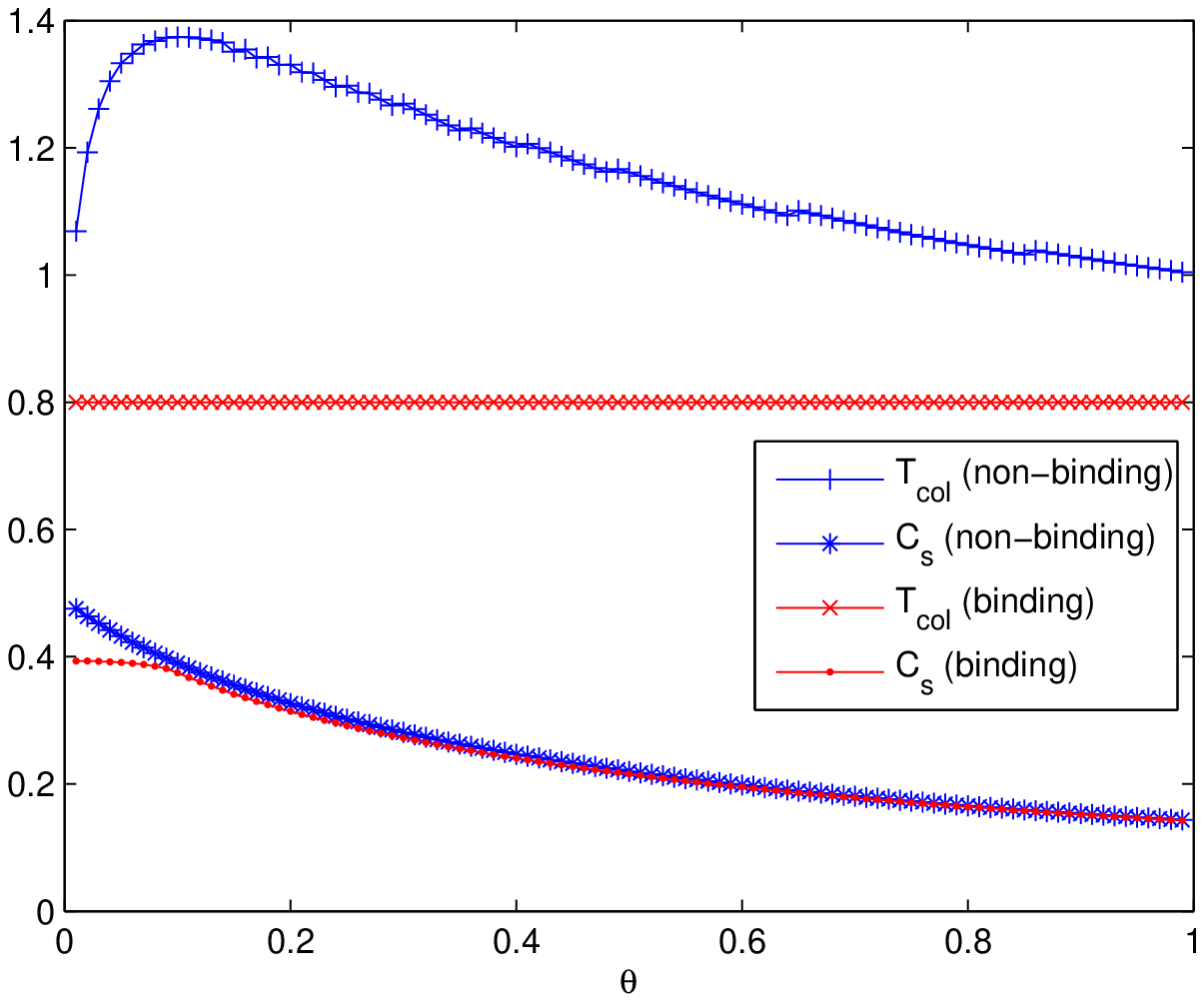}}%
\caption{Solution to the protocol design problem
for $\theta$ between $0.01$ and $0.99$ when $N = 10$, $T_{int} = 100$, and $T_{pac} = 50$:
\protect\subref{fig:tunc-a} optimal protocols, and
\protect\subref{fig:tunc-b} the values of $T_{col}$ and $C_s$ at the optimal protocols.}
\label{fig:tunc}%
\end{figure}

We investigate the impact of the fairness level
on optimal protocols and their performance. We first
consider sufficiently large $\gamma$ so that the constraint is non-binding. Fig.~\ref{fig:tunc}\subref{fig:tunc-a} shows optimal
protocols $(q^*,r^*)$ as $\theta$ varies from $0.01$ to $0.99$ when the constraint is non-binding.
As $\theta$ increases, $q^*$ stabilizes around $0.10$
quickly whereas $r^*$ keeps increasing but at a diminishing rate.
As $\theta$ is larger, contention periods occur more frequently during an \emph{off} period,
and thus it becomes more important to resolve contention among the secondary users quickly
by having $r \approx 1/2$ when maximizing $C_s$.
Fig.~\ref{fig:tunc}\subref{fig:tunc-b} plots the values of $T_{col}$ and $C_s$
at $(q^*,r^*)$. As $\theta$ increases, $T_{col}$ increases, reaches a peak at $\theta = 0.1$,
and then decreases, whereas $C_s$ decreases monotonically. The negative relationship
between $C_s$ and $\theta$ can be interpreted as a trade-off
between channel utilization and short-term fairness.

Since $T_{col}$ at $(q^*,r^*)$ ranges between $1.00$
and $1.37$, we set $\gamma = 0.8$ to analyze the protocol design
problem with a binding constraint.
Fig.~\ref{fig:tunc}\subref{fig:tunc-a} shows optimal
protocols $(q^o,r^o)$ with $\gamma = 0.8$ while Fig.~\ref{fig:tunc}\subref{fig:tunc-b} plots
the values of $T_{col}$ and $C_s$ at $(q^o,r^o)$, as $\theta$ varies from $0.01$ to $0.99$. Note that the optimal protocols are at the corner
with $r^o = \epsilon$ for $\theta \leq 0.09$. Imposing the constraint $T_{col} \leq 0.8$
limits the values of $q$ and $r$. The differences between $q^*$ and $q^o$ and between $r^*$ and $r^o$
are larger for smaller $\theta$ in the region $[0.1, 1]$ because requiring $T_{col} \leq 0.8$ imposes a stronger constraint
for smaller $\theta$, which can be seen by comparing the values of $T_{col}$
with binding and non-binding constraints in that region. The impact of the constraint on $C_s$ is marginal as long
as the optimal protocols are interior.

\subsection{Estimated Number of Secondary Users}

\begin{figure}%
\centering
\includegraphics[width=0.5\textwidth]{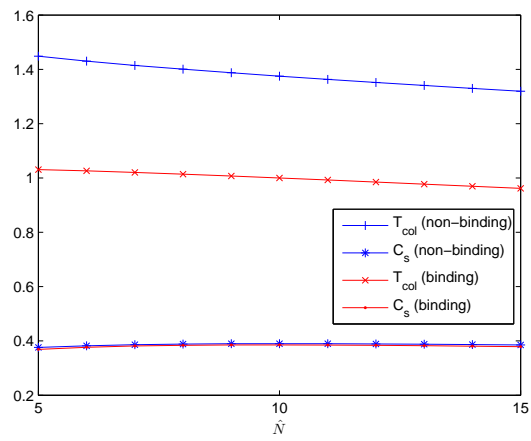}
\caption{Values of $T_{col}$ and $C_s$
for $\hat{N}$ between 5 and 15
when $N = 10$, $T_{int} = 100$, $T_{pac} = 50$, and $\theta = 0.1$.}
\label{fig:esti}%
\end{figure}

Suppose that the protocol designer solves the protocol design problem
for each possible $N$ and prescribes the obtained protocols for the secondary users as a function of $N$.
If the secondary users know the exact number of secondary users sharing the channel,
an optimal protocol can be implemented. Here we consider a scenario where
the secondary users choose an optimal protocol based on their (possibly incorrect)
estimates of the number of secondary users.
For simplicity, we assume that all the secondary users have the same estimate. We consider $N = 10$ and
the estimated number of secondary users, denoted by $\hat{N}$, between $5$ and $15$.
In Fig.~\ref{fig:esti}, we plot the values of $T_{col}$ and $C_s$
when the $N$ secondary users follow the optimal protocol computed assuming $\hat{N}$ secondary users.
As before, we consider the two cases of non-binding and binding constraints, with $\gamma = 1$ for the binding
constraint. In both cases, optimal $q$ decreases with the estimated number of secondary users
while optimal $r$ is almost constant, as shown in Fig.~\ref{fig:nunc}\subref{fig:nunc-a}.
The overall interference level from the secondary users reduces as $\hat{N}$ increases,
and thus $T_{col}$ decreases with $\hat{N}$.
$C_s$ is not affected much by $\hat{N}$, reaching a peak when $\hat{N} = N$.
This result suggests that channel utilization
is robust to errors in the estimation of the number of secondary users.
Note that, in the case of the binding constraint, the constraint is violated slightly when an
underestimation occurs, i.e., $\hat{N} < N$.
In order to offset this effect, the protocol designer can choose an estimation procedure
that is biased toward overestimation, or specify a smaller $\gamma$ than the required threshold.

\section{Enhancement Using Longer Memory}

We have adopted protocols with one-slot memory for their simplicity.
Protocols with one-slot memory not only are easy to design and implement
but also allow us to use Markov chains to study performance.
However, as illustrated in \cite{park}, it is possible to obtain performance improvement
by utilizing longer memory. In this section, we explain how longer memory can
help reduce the average number of collisions
and bound the maximum number of collisions experienced by the primary user in an \emph{on} period.
Let $p_i^{\tau}$ be the transmission probability of secondary user $i$ in slot $\tau$.
A protocol with $B$-slot memory that enhances a $\theta$-fair non-intrusive protocol $f$
can be expressed as follows:
\begin{description}
\item[(P1)] If $y_i^{t-2} = success$ and $y_i^{t-1} = failure$, then $p_i^t = 0$.

\item[(P2)] If $y_i^{t-B} = \cdots = y_i^{t-1} = failure$, then $p_i^t = 0$.

\item[(P3)] Otherwise, $p_i^t = f(y_i^{t-1})$.
\end{description}

\subsection{Reducing the Average Number of Collisions Experienced by the Primary User}

(P1) requires that a secondary user that experiences a collision following a success back off.
Note that a collision following a success cannot occur in an \emph{off} period by non-intrusiveness,
and thus (P1) does not affect performance in an \emph{off} period.
The only possible occasion in which a collision follows a success is when the
primary user starts transmitting. Therefore, if a secondary user experiences
a collision following a success, it can infer than an \emph{on} period has started.
According to a $\theta$-fair non-intrusive protocol, a secondary user transmits
with probability $r$ after a collision, which yields $d(1) = (1-\theta)
[ (\mathbf{I} - \mathbf{Q}_{\textrm{\emph{on}}})^{-1} \mathbf{e} ]_1$ in \eqref{eq:d1}.
By imposing (P1), we can reduce the value to $d(1) = 1 - \theta$, which in turn
reduces the value of $T_{col}$. For example, with $N = 10$, $T_{int} = 100$, $T_{pac} = 50$, $\theta = 0.1$,
and $(q,r) = (q^*,r^*) = (0.10, 0.37)$,
(P1) reduces $d(1)$ from $1.426$ to $0.9$ and $T_{col}$ from $1.376$ to $0.954$.

\subsection{Bounding the Maximum Number of Collisions Experienced by the Primary User}

In the range of parameter values considered in Section V, the average number of
collisions experienced by the primary user in an \emph{on} period is reasonably small,
not exceeding 1.5 slots, even without a constraint imposed on it. However, as colliding secondary users transmit with probability $r > 0$,
the realized number of collisions in an \emph{on} period can be arbitrarily large
with positive probability. That is, the worst-case number of collisions in an \emph{on} period
is unbounded under a $\theta$-fair non-intrusive protocol. We can bound the maximum
number of collisions in an \emph{on} period by imposing (P2), which requires
a secondary user that experiences $B$ consecutive collisions to back off.
Since non-colliding secondary users wait after a collision, colliding secondary users must have the same number
of consecutive collisions in any slot. Thus, secondary users experiencing $B$ consecutive collisions
back off simultaneously, yielding a slot that can be utilized by the primary user.
Therefore, the primary user cannot experience more than $B$ collisions
in an \emph{on} period. When $B$ is chosen moderately large, $B$ consecutive collisions
rarely occur in an \emph{off} period, and thus (P2) has a negligible impact on
the success probability of the secondary users $P_s$ while it reduces $T_{col}$.
(P2) can be considered
as a safety device to limit the number of collisions that the primary user
can experience during an \emph{on} period.

\section{Conclusion}

In this paper, we have considered a scenario in which a primary user shares
a channel with secondary users that cannot distinguish the signals of the primary user
from those of a secondary user.
We have shown that a class of distributed MAC protocols can
coordinate access among the secondary users while restricting interference
to the primary user, thereby overcoming the limited sensing ability of the secondary users
at the PHY layer. The basic ideas underlying the proposed protocols
can be exploited in different settings. For example,
in a random access network with CSMA/CA, protocols with memory can be used to
adjust the back-off parameters of secondary users based on their own transmission results
and obtained channel information. Also, we can provide quality-of-service
differentiation to secondary users by specifying different protocol parameters across
secondary users.
The fairness level for a secondary user determines the average number
of its consecutive successes, while the transmission probabilities following
an idle or a collision slot determine the probability that a secondary user is chosen as
the successful user for the next success period in a contention period.
Finally, the enhanced protocols with longer memory suggest the potential of
observed patterns in history as a substitute for explicit information passing.
As users make decisions based on history under a protocol with memory,
users can adjust their behavior to the network environment or the states of other users
by extracting information from history.



\vfill

\end{document}